\documentclass[prl,twocolumn,showpacs,superscriptaddress]{revtex4}
\usepackage[dvips]{graphicx}
\usepackage{dcolumn}
\usepackage{bm}
\usepackage{amssymb}

\DeclareGraphicsExtensions{eps}

\begin{document}
\preprint{APS/123-QED}
\title{Ground state of the Kagom\'e-like S=1/2 antiferromagnet, Volborthite~Cu$_3$V$_2$O$_7$(OH)$_2 \centerdot 2$H$_2$O}
\author{F. Bert}
\author{D. Bono}
\author{P. Mendels}
\affiliation{%
Laboratoire de Physique des Solides, UMR 8502, Universit\'{e}
Paris-Sud, 91405 Orsay, France.}%
\author{F. Ladieu}
\affiliation{Service de Physique de l'\'Etat
Condens\'{e}, DSM, CEA Saclay,
91191 Gif-sur-Yvette Cedex, France.}
\author{F. Duc}
\author{J.-C. Trombe}
\author{P. Millet}
\affiliation{Centre d'\'Elaboration des Mat\'eriaux et d'\'Etudes
Structurales, CNRS UPR 8011, 31055 Toulouse, France.}

\date{\today}

\begin{abstract}
Volborthite compound is one of the very few realizations of S=1/2
quantum spins on a highly frustrated kagom\'e-like lattice. Low-T
SQUID measurements reveal a broad magnetic transition below 2~K
which is further confirmed by a peak in the $^{51}$V nuclear spin
relaxation rate ($1/T_1$) at 1.4~K$\pm$0.2~K. Through $^{51}$V
NMR, the ground state (GS) appears to be a mixture of different
spin configurations, among which 20\% correspond to a well defined
short range order, possibly of the $\sqrt{3} \times \sqrt{3}$
type. While the freezing involve all the Cu$^{2+}$ spins, only
40\% of the copper moment is actually frozen which suggests that
quantum fluctuations strongly renormalize the GS.

\end{abstract}

\pacs{75.30.Cr,75.50.Lk,76.60.-k}
\maketitle

In the quest for novel states of condensed matter, frustration has
emerged as a key concept~\cite{Ramirez01}. In magnetic systems,
competitive interactions resulting from the geometry of the
lattice, especially for corner sharing networks, can lead to a
macroscopic entropy at T=0~K and could favor a novel spin liquid
state. Theoretically, Heisenberg S=1/2 antiferromagnet on a 2D
kagome lattice is predicted to lead to such an exotic quantum
state~\cite{Misguich03,mila98}. Actually very few real systems
approach this model. So far, the Cr$^{3+}$ (S=3/2) kagom\'e
bilayer compounds SrCr$_{9p}$Ga$_{12-9p}$O$_{19}$ (SCGO) and
Ba$_2$Sn$_2$ZnGa$_{10-7p}$Cr$_{7p}$O$_{22}$ ($p<0.97$) have been
the experimental archetypes of Heisenberg frustrated
antiferromagnets. In this context, the recent rediscovery of
Volborthite Cu$_3$V$_2$O$_7$(OH)$_2 \centerdot
2$H$_2$O~\cite{Lafontaine90,Hiroi01} is a major step towards a
realization of the theoretical model. The magnetic lattice of this
natural antiferromagnet consists of quantum S=1/2 (Cu$^{2+}$)
spins sitting at the vertices of well separated ($7.2 \AA$)
kagome-like planes. The lattice displays a monoclinic distortion,
possibly yielding two Cu-Cu interaction constants $J_1 \neq J_2$,
which does not seem to impact on the characteristic fingerprints
of frustration. Indeed, despite strong antiferromagnetic
interactions ($J \simeq 90$~K), no transition towards an ordered
state has been detected down to 1.8~K, neither in susceptibility
nor in heat capacity measurements~\cite{Hiroi01}. Instead, a
maximum in both these quantities is observed around 20~K, probably
reflecting the enhancement of short range correlations, and
defines a new low energy scale as a result of
frustration~\cite{Mendels00,Bono04}. Besides, as in kagom\'e
bilayers~\cite{Uemura94}, muon spin relaxation ($\mu$SR)
experiments detected no sign of static spin freezing but rather
temperature independent spin fluctuations persisting down to
50~mK, indicative of a fluctuating quantum GS~\cite{Fukaya03}.
Only in one ESR study the existence of an internal field was
evidenced at 1.8~K which was interpreted as short range
order~\cite{Okubo01}.

In this Letter, we show, for the first time, that volborthite
undergoes, around 1.3~K, a transition to a frozen state which we
characterize using $^{51}$V NMR as a local probe of magnetism.
Note that, so far, NMR investigation of the GS in the kagom\'e
bilayers had proved to be impossible because of a wipe out of the
NMR intensity at low T~\cite{Limot02}. The volborthite powder
samples were prepared by refluxing an aqueous suspension of
V$_2$O$_5$ and a basic copper (II) carbonate salt Cu(OH)2-Cu(CO3)
for several days. No spurious phase was detected in X-ray
measurements. SQUID and NMR measurements above 2~K and $\mu$SR
experiments on these samples~\cite{Bert04} were similar to the
above described published data.

\begin{figure}
\includegraphics[width=7.1cm,height=8cm]{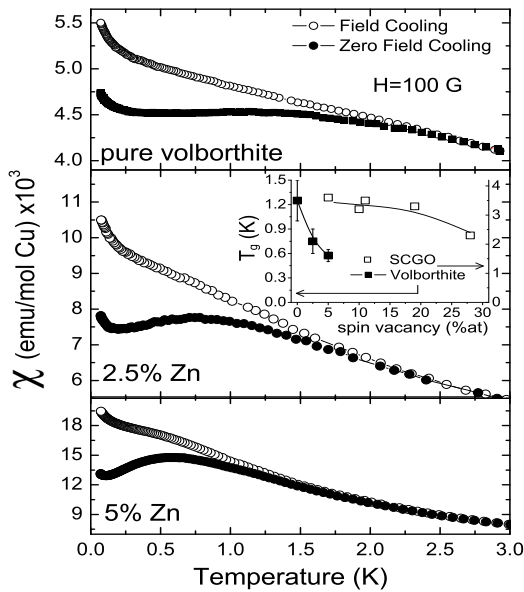}
\caption{\label{squid} SQUID dc susceptibility in
Cu$_{3(1-x)}$Zn$_{3x}$V$_2$O$_7$(OH)$_2 \centerdot 2$H$_2$O
measured after two cooling protocols, with (open symbols) and
without (full symbols) a 100~Oe applied field. Inset: comparison
of $T_g$ versus S=0 impurity dilution rate in volborthite and SCGO
bilayer compounds.}
\end{figure}

The low T static susceptibility of volborthite is presented in
Fig.~\ref{squid} (top panel). Below 2~K, the separation between
the field cooled (FC) and zero field cooled (ZFC) susceptibilities
reveals a spin freezing process. At 1.2~K, "ageing"
effects~\cite{Vincent97} have been observed (not shown here) which
ascertains the glassy nature of the GS. At variance with textbook
spin glass transition, the FC susceptibility does not level off
below the transition and the hardly detectable broad maximum on
the ZFC branch occurs below the FC/ZFC separation~\footnote{The
up-turn at $T < 0.1$~K, is due to few ppm magnetic impurities in
the grease used to ensure thermalization.}. This suggests a rather
broad distribution of freezing temperatures as, for instance, in
super-paramagnets where spin clusters get progressively blocked on
lowering T.

To address the important issue of the intrinsic character of this
glassy low-T phase, the susceptibility of two zinc substituted
samples is also reproduced in Fig.~\ref{squid}. Zn atoms enter
Volborhtite lattice without noticeable distortion. Since Zn$^{2+}$
is non-magnetic, the Zn/Cu substitution results in diluting the
kagome magnetic lattice. The freezing temperature $T_g$, defined
here as the temperature of the ZFC maxima, are plotted in the
inset of Fig.~\ref{squid} as a function of the magnetic lattice
dilution and compared to the same data for SCGO
compounds~\cite{Limot02}. In marked contrast with SCGO, $T_g$ for
Volborthite is strongly affected by dilution. This reminds us of
the drastic reduction upon dilution of the dynamical plateau value
seen in $\mu$SR experiments~\cite{Bert04} and suggests that the
bilayer topology likely better accommodates defects than the
single layer one in volborthite. Nonetheless, in both systems, the
random dilution of the magnetic network reduces $T_g$ and hence
simple magnetic dilution as a source of disorder cannot explain,
alone, the spin-glass like transition.

Microscopic insight into this ground state is provided by our
low-T $^{51}$V NMR experiments. The non-magnetic V$^{5+}$ ions are
situated above or below the center of the stars which constitutes
the kagom\'e lattice. They thus probe symmetrically the magnetism
of six Cu$^{2+}$ ions belonging to the same hexagon through an
hyperfine coupling, estimated to be $A \simeq 7.7$~kOe from
susceptibility versus NMR shift measurements (see
Ref.~\cite{Bert04}) above 90~K in agreement with the value
estimated in Ref.~\cite{Hiroi01}.

\begin{figure}
\includegraphics[scale=0.8]{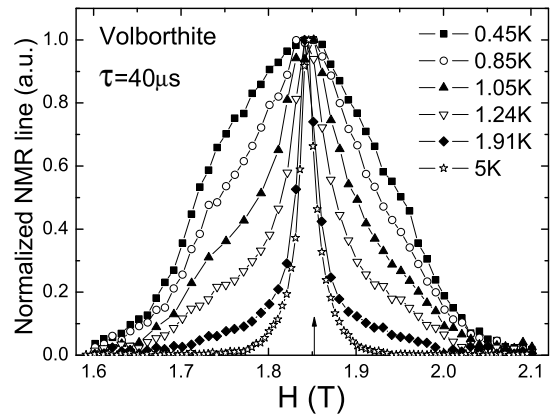}
\caption{\label{RMNLT} $^{51}$V NMR spectra measured at 20.733~MHz
as a function of the applied field. For clarity, the spectra have
been normalized to their maximum value. The vertical arrow shows
the position $H_0$ of the unshifted reference line.}
\end{figure}

\begin{figure}[t]
\includegraphics[scale=0.7]{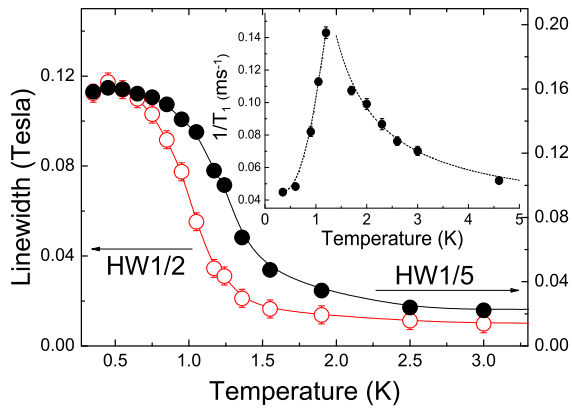}
\caption{\label{T1} Half widths of the NMR spectra
measured at 1/2 (HW1/2, open symbols) and 1/5 (HW1/5, full
symbols) of the maximum height. Inset : $T$-dependence of the
$^{51}$V spin-lattice relaxation rate (1/T$_1$) measured at the
maximum intensity of the spectra (H=1.85~T). Lines are guides for
the eye.}
\end{figure}

The nuclear spin lattice relaxation has been measured by the
saturation-recovery method and fitted to a sum, with T-independent
coefficients, of four exponential terms with relaxation rates
proportional to $1/T_1$ as expected for a S=7/2 nuclear
spin~\cite{Narath67} in the case of partial saturation of the NMR
line. The divergence of $1/T_1$ at 1.4~K$\pm$0.2~K (Fig.~\ref{T1},
inset) is a strong evidence that the vanadium probe indeed feels a
transition in agreement with our susceptibility measurements.
Following the analysis of Ref.~\cite{Fukaya03} of the zero field
$\mu$SR relaxation rate below 1~K ($\lambda=5~\mu$s$^{-1}$ in our
sample~\cite{Bert04}), we get 3.5~MHz for the copper spin
fluctuation rate. From the hyperfine coupling $A$, we would then
expect the corresponding $^{51}$V $1/T_1$ to be $\simeq
20$~ms$^{-1}$, much higher than the value actually measured. This
suggests that spin fluctuations, seen in $\mu$SR, are efficiently
filtered by the nuclear probe. We shall come back to this point
later.

Characteristic $^{51}$V ($\gamma / 2 \pi=11.1923$~MHz/T) NMR
spectra measured at $\nu_0=20.733$~MHz, below 5~K, are presented
in Fig.~\ref{RMNLT}. At these low T, the quadrupolar splitting of
the S=7/2 vanadium NMR line is masked by a large magnetic
broadening which reflects the width of the field distribution at
the vanadium site. On lowering T, the NMR line broadens
drastically around 1.5~K and then saturates below 0.6~K. At a
lower 12.548~MHz irradiating frequency, we checked that this
saturation is field independent, and therefore reflects, as
expected, a frozen field distribution in the GS. Upon closer
inspection of the T-dependence of the lineshape, we note first the
appearance of a broad background feature below 1.5~K. In order to
track qualitatively this broad component we chose rather
arbitrarily to plot the half width at 1/5$^{th}$ of the maximum in
Fig.~\ref{T1}. Then the main central line, which width is roughly
the width at half maximum of the spectra (open symbols in
Fig.~\ref{T1}), starts to broaden rapidly and below 1~K, its
lineshape changes from lorentzian-like to gaussian-like. The two
distinct features, broad background and central line, in the low-T
spectra suggest that the NMR line, below 1.5~K, is no more
homogenous but results from at least two different types of
magnetic environments of vanadium. This is probably related, as
well, to the progressive freezing observed in macroscopic
susceptibility. It is an important finding of this study as it
demands a special magnetic ordering in the GS leading to two
different V sites.

\begin{figure}
\includegraphics[width=6.6cm,height=8cm]{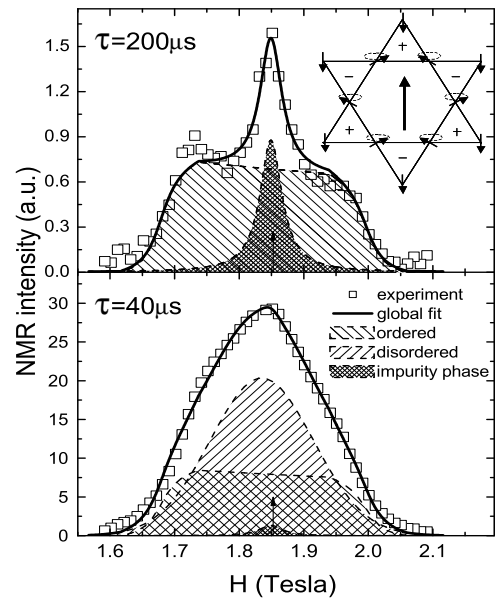}
\caption{\label{contrast} $^{51}$V NMR spectra measured for two
different delays $\tau$ at T=0.35~K. Note the different scales in
the top and bottom panel. The solid lines are fits to the data and
result from the sum of three components (dashed, dashed-dotted and
dotted lines) as explained in the text. Inset: scheme of the
proposed $\sqrt 3 \times \sqrt 3$ local order responsible for the
ordered component (dashed line). The vector at the center of the
hexagon stands for the resulting frozen field at the vanadium
site.}
\end{figure}

We now focus on the T=0.35~K spectra (Fig.~\ref{contrast}) in the
well established GS of volborthite. By comparison of the
integrated intensity with $T>2~K$ data, we checked that all sites
are detected at this low-T and we therefore probe the bulk
properties of the sample. The top and bottom spectra have been
obtained in the same conditions except for a different duration
$\tau$ in the pulse sequence $\pi/2 - \tau - \pi/2$. This contrast
procedure allows us to separate the two components of the
spectrum. They indeed prove to have different spin-spin relaxation
times $T_2$ which we determined by standard $T_2$ measurements at
$H=1.845$~T and $H=1.955$~T.

 In the long time
spectrum (top panel), the slowly relaxing ($T_2=105~\mu$s) broad
background clearly stands out, without the fast relaxing
($T_2=60~\mu$s) gaussian-like component and appears to be
rectangular shaped. An unshifted narrower line also appears on
this spectrum. However, because of its much longer $T_2=240~\mu$s,
it eventually represents less than 1$\%$ of the total sample and
probably arises from some small impurity phase. The rectangular
lineshape is a clear signature of the presence of one well defined
frozen field $H_f$ at the vanadium site arising from the
neighboring copper moments. The resonance condition writes
$H_0=2\pi \nu_0 / \gamma=\| \mathbf{H} + \mathbf{H}_f \|$ where
$H$ is the applied field. Due to powder distribution,
$\mathbf{H}_f$ is randomly oriented with respect to $\mathbf{H}$.
In the limit $H_f \ll H_0$ one would expect a rectangular
lineshape with cut-offs at $H_0 \pm H_f$. An exact derivation of
the NMR line $P(H)$ yields
\begin{equation}
P(H)=(H_0^2+H^2-H_f^2)/(4 H_f H^2)
\end{equation}
for $|H-H_0|< H_f$. The component labelled "ordered" in
Fig.~\ref{contrast} is a fit with such a model with $H_f=0.16$~T
and a narrow distribution (0.02~T HWHM) of this frozen field. The
solid line in the top panel is a global fit to the NMR line,
including the impurity phase. In order to produce a well-defined
amplitude of the frozen field at the vanadium site, short-range
order of the six neighboring Cu$^{2+}$ moments must exist.
 In a classical picture,
the energy is minimized on the kagom\'e lattice, when the spins
are at 120$^\circ$ from each other on each triangle. Following
most of the theoretical works~\cite{Reimers93}, we assume a $\sqrt
3 \times \sqrt 3$ type short range order, sketched in the inset of
Fig.~\ref{contrast}, which corresponds to alternating chiralities
(+ and - signs) on neighboring triangles.
 Such a short range order indeed leads to a well defined $H_f=3H_{Cu}$
 resulting field at the vanadium site, where $H_{Cu}$ is the frozen field arising from each Cu$^{2+}$.
On the contrary, a uniform chirality order ($\mathbf{q}=0$)
would lead to a null field at the vanadium site which cannot
reproduce our NMR data. In addition, the classical $\sqrt 3 \times
\sqrt 3$ order favors local collective excitations made of a
coherent out-of-plane rotation of the six spins belonging to a
same hexagon as depicted in the inset of Fig.~\ref{contrast}. It
is remarkable that such excitations, which cost zero energy, do
not affect the resulting field at the vanadium site, neither by
hyperfine nor dipolar coupling, and are, therefore, "filtered" out
at the symmetric vanadium site. This could explain why we measured
a small $1/T_1$ which, besides, decreases at low temperature, as
in usual static phases, whereas, in $\mu$SR experiments, muons
which sit in less symmetric positions, still feel a strong
dynamics which persists below $T_g$~\cite{Fukaya03,Bert04}. From
$H_f$ and the hyperfine coupling constant, we get $0.41 \mu_B$ for
the frozen Cu$^{2+}$ moment contributing to this slow relaxing
component. This small value, compared to 1$\mu_B$ expected for a
spin 1/2, demonstrates that zero point quantum fluctuations
strongly affect the volborthite GS.

 In the bottom panel, combining the formerly discussed slow relaxing components
 with a frozen gaussian-like component ("disordered" label) results in the solid line which reproduces well the
 short time experimental spectra. Taking into account the different
 $T_2$ values, we evaluate the corresponding sample fractions, in the limit $\tau \rightarrow 0$, to
 be $\simeq$80\% for the gaussian like component and
 $\simeq$20\% for the rectangular-shaped one.
A gaussian-like frozen component is easily obtained provided that
the Cu$^{2+}$ frozen moments belonging to a same hexagon are
randomly oriented as in the case of a spin glass. In this random
picture, we extract also a small $0.44\mu_B$ value for the frozen
copper moment consistent with the previous one. A gaussian like
component could also come from several frozen field values,
smaller than in the $\sqrt 3 \times \sqrt 3$ case and slightly
distributed, arising from various other spin configurations. A
crude illustration of this scenario can be obtained if one assumes
a random distribution of the chiralities on the kagom\'e lattice.
One finds then that 4\% of the hexagons are in the $\sqrt 3 \times
\sqrt 3$ configurations while the other configurations lead to
either $H_f=0$ (54\%) or $H_f=\sqrt 3 H_{Cu}$ (41\%). Our
experimental value already indicates that, within this classical
framework, the $\sqrt 3 \times \sqrt 3$ configuration has to be
favored by some mechanism. An exact quantum calculation of the
possible states in a finite kagom\'e spin 1/2 cluster would
certainly allow a better quantitative understanding of this
gaussian-like component in this scenario.
 At the nanoscopic scale of the NMR probe, it is difficult to
decide whether these two vanadium sites in the GS reveal different
domains or appear as a mixture in a single phase. However, both
configurations freeze at approximatively the same temperature and,
for both, we found a similar frozen fraction of the
Cu$^{2+}$ moments. Both arguments favor an original single phase
description.

The disorder of the GS is a challenge to our understanding of
highly frustrated magnets, given that volborthite is probably the
purest model system known so far. From a comparison with Zn
diluted samples, we estimated an upper limit of 1\% spin vacancy
like defects in the pure volborthite sample. Theoretical
works~\cite{Chandra93,Ferrero03} have put forward the possibility
of an intrinsic "topological" spin glass state arising from
frustration alone contrary to usual spin glasses where both
frustration and disorder are responsible for glassiness. Alternatively, in
Ref.~\cite{Bono04b}, it was argued that a small amount of disorder could
induce a spin glass like state provided that a coherent RVB background couples
very efficiently the defect centers. This scenario
 naturally explains the simultaneous occurrence of the glassy-like
transition and the dynamical plateau in $\mu$SR experiments and is
consistent with our finding of a two component NMR signal, arising
from vanadium nuclei far and close to a defect.

In conclusion , volborthite which presents all the well
established signatures of a spin liquid, namely no freezing at $T
\simeq \theta_{CW}$ and a dynamical plateau at $T \rightarrow 0$,
allows, for the first time, a detailed NMR local investigation of
the GS. This enabled us to study the internal field configurations
and their dynamics. Further, this opens new avenues for refined
theoretical calculations which are necessary to reveal
the influence of the possible dissymmetry of the interactions and
the actual texture of the GS.



\end{document}